\documentclass[twocolumn,aps,pra,superscriptaddress,showpacs,tightenlines]{revtex4}
\usepackage{amsmath}
\usepackage{amsfonts}
\usepackage{graphicx}
\usepackage{epsfig}
\usepackage{color}
\usepackage{hyperref}
\hypersetup{
    colorlinks=true, 
    linktoc=all,     
    linkcolor=blue,  
    citecolor=blue}

\begin{document}

\title{Quantum coherence in ultrastrong optomechanics}
\author{ Dan Hu}
\affiliation{School of Natural Sciences, University of California, Merced, CA 95343, USA}
\author{Shang-Yu Huang}
\affiliation{Department of Physics and Center for Theoretical Sciences, National Taiwan University, Taipei 10617, Taiwan}
\author{Jie-Qiao Liao}
\affiliation{CEMS, RIKEN, Saitama 351-0198, Japan}
\author{Lin Tian}
\email{ltian@ucmerced.edu}
\affiliation{School of Natural Sciences, University of California, Merced, CA 95343, USA}
\author{Hsi-Sheng Goan}
\email{goan@phys.ntu.edu.tw}
\affiliation{Department of Physics and Center for Theoretical Sciences, National Taiwan University, Taipei 10617, Taiwan}
\affiliation{Center for Quantum Science and Engineering and National Center for Theoretical Sciences, National Taiwan University, Taipei 10617, Taiwan}

\begin{abstract} 
Ultrastrong light-matter interaction in an optomechanical system can result in nonlinear optical effects such as photon blockade. The system-bath couplings in such systems play an essential role in observing these effects. Here we study the quantum coherence of an optomechanical system with a dressed-state master equation approach. Our master equation includes photon-number-dependent terms that induce dephasing in this system. Cavity dephasing, second-order photon correlation, and two-cavity entanglement are studied with the dressed-state master equation. 
\end{abstract}
\maketitle

\section{Introduction~\label{sec1}} 
Cavity optomechanics studies quantum effects induced by light-matter interaction between cavity and mechanical modes~\cite{Aspelmeyer2013,Chen2013}. Many such effects have been realized in recent experiments, including the preparation of quantum ground state, the observation of strong optomechanical coupling, and the coherent conversion of photon states via a mechanical interface~\cite{Groundstate1, strongcouplingExp2, strongcouplingExp3, Groundstate5, Groundstate6, strongcouplingExp4, strongcouplingExp5, Appl3, Appl4}. Among recent theoretical works, studies of optomechanical systems in the single-photon strong or ultrastrong coupling regime have predicted many interesting nonlinear optical effects such as photon blockade, phonon sidebands and nonlinear optomechanically-induced transparency~\cite{Ultrastrong1, Ultrastrong2, Ultrastrong3, Liao2012, Kronwald2013, Liao2013, Komar2013, Nunnenkamp2012, Xu2013A, Kronwald2013B, Qian2012, AkramNJP2013}. With the strength of the single-photon optomechanical coupling comparable to the mechanical frequency and the cavity bandwidth, the optomechanical systems can demonstrate strong nonlinearity. It is promising to reach this regime in several experimental systems~\cite{strongcouplingExp2, strongcouplingExp3, Groundstate5, Groundstate6, strongcouplingExp4, strongcouplingExp5, UltrastrongExp1, UltrastrongExp2}. In addition, recent theoretical works have shown that ultrastrong coupling could be achieved by various quantum engineering schemes~\cite{Xuereb2012, Sillanpaa2013, Rimberg2013, Lu2013, Liao2014}.

The cavity and the mechanical modes in an optomechanical system are subject to environmental noise, which causes decoherence and plays a crucial role in studying the nonlinear optical effects. The system-bath couplings can be treated with a master equation approach. Very often, a standard master equation (SME) is used to describe the damping and thermal excitations. For example, the contributions of the mechanical bath can be in the form of $\mathcal{D}[\hat{b}]\rho(t)$ and $\mathcal{D}[\hat{b}^{\dag}]\rho(t)$, where $\hat{b}$ is the annihilation operator of the mechanical mode, $\mathcal{D}[\hat{o}]\rho(t)=\frac{1}{2}[2\hat{o} \rho(t)\hat{o}^\dagger -\hat{o}^\dagger\hat{o}\rho(t)-\rho(t)\hat{o}^\dagger\hat{o}]$ is the Lindblad superoperator for operator $\hat{o}$, and $\rho(t)$ is the density matrix of the optomechanical system at time $t$. Such treatment is based on the assumption that the optomechanical coupling is much weaker than the mechanical frequency, and thus does not seriously modify the eigenstates of this system. Under this assumption, each system mode is only affected by their corresponding bath modes. However, in the single-photon strong or ultrastrong coupling regime, photons in the eigenstates are strongly dressed by phonon excitations of the mechanical mode, and this assumption is not valid anymore~\cite{qopticsbook, dePonte2004, BLHu2008, TianPRB2011}. 

Here we study the quantum coherence and dynamics of an optomechanical system in the ultrastrong coupling regime with an appropriate master equation approach. In our method, we decompose the system operators in terms of the eigenstates (dressed states) of the optomechanical system and derive the master equation under this decomposition. This approach was previously used to study strongly-coupled harmonic oscillators with linear coupling~\cite{dePonte2004, BLHu2008} and a mechanical resonator coupled to a two-level-system defect~\cite{TianPRB2011}. Our master equation contains photon-number-dependent terms in the form of $\mathcal{D}[\hat{b}-\beta_0\hat{N}_c]\rho(t)$, $\mathcal{D}[\hat{b}^{\dag}-\beta_0\hat{N}_c]\rho(t)$ and $\mathcal{D}[\hat{N}_c]\rho(t)$, which cause mechanical damping as well as cavity dephasing. Counter-intuitively, the term $\mathcal{D}[\hat{N}_c]\rho(t)$ that generates dephasing between different photon Fock states is not induced by cavity bath modes. It is originated from mechanical bath modes that influence the state of the cavity via light-matter interaction. We show that at high temperature our master equation generates faster cavity dephasing and entanglement decay when compared with the SME. The second-order photon correlation given by our master equation also demonstrates more classical behavior than that of the SME at high temperature, predicting photon bunching in some regions of photon antibunching predicted by the SME. Our results indicate that the coherence of an optomechanical system could be strongly influenced by the ultrastrong coupling, and the SME may not be sufficient for studying this system. 

This paper is organized as follows. In Sec.~\ref{sec2}, we present the master equation derived in the dressed-state basis of an optomechanical system and compare this master equation with the SME. We then study the quantum coherence properties of an optomechanical system governed by this master equation in Sec.~\ref{sec3}, Sec.~\ref{sec4} and Sec.~\ref{sec5}, respectively, on the dephasing of the cavity state, the second-order photon correlation in the stationary state of the cavity, and the dynamics of two-cavity entanglement. Conclusions are given in Sec.~\ref{sec6}.

\section{Dressed-state master equation\label{sec2}} 
We consider an optomechanical system with one cavity mode and one mechanical mode coupling via
radiation-pressure interaction. The Hamiltonian of this system is ($\hbar=1$)
\begin{equation}
\hat{H}_{s}=\omega_c\hat{a}^{\dagger}\hat{a}+\omega_m\hat{b}^{\dagger}\hat{b}-g_0\hat{a}^{\dagger}\hat{a}(\hat{b}+\hat{b}^{\dagger}), \label{eq:Hs}
\end{equation}
where $\omega_c$ ($\omega_m$) is the cavity (mechanical) frequency, $g_0$ is the strength of the single-photon optomechanical coupling, $\hat{a}$ ($\hat{b}$) is the annihilation operator of the cavity (mechanical) mode. The eigenstates of this coupled system can be written as
\begin{equation}
|n,k^{(n)}\rangle=|n\rangle_{c}\otimes e^{n\beta_0(\hat{b}^\dagger-\hat{b})}|k\rangle_{m} \label{eigstate}
\end{equation}
with cavity photon number $n$ and phonon number $k$ for the mechanical mode. Here the state $|k^{(n)}\rangle$ is the mechanical Fock state $|k\rangle_{m}$ shifted with a displacement $n\beta_0$ that is proportional to the cavity photon number $n$ and $\beta_0=g_0/\omega_m$. In other words, the eigenstates are dressed states in which the cavity photon excites a photon-number-dependent mechanical displacement due to the optomechanical coupling. The corresponding eigenenergies of these states are $\varepsilon_{n,k}=n\omega_c+k \omega_m-n^2 g_0^2/\omega_m$. In this work, we study the optomechanical system in the ultrastrong coupling regime with the single-photon optomechanical coupling $g_{0}$ comparable to (or larger than) the mechanical frequency and the cavity bandwidth $\kappa$. In this regime, the mechanical component of the eigenstates is strongly shifted by the optomechanical coupling with a displacement proportional to the cavity photon number~\cite{Ultrastrong1, Ultrastrong2, Ultrastrong3, Rimberg2013}. 

The cavity and the mechanical modes couple to environmental degrees of freedom which induce damping and thermal excitations in the optomechanical system. The system-bath couplings can be written as $\hat{H}_{b}^{I}=\hat{H}_{cb}^{I}+\hat{H}_{mb}^{I}$ in the interaction picture with~\cite{qopticsbook}
\begin{eqnarray}
\hat{H}_{cb}^{I}&=&\hat{a}^{\dagger}(t) \hat{\Gamma}_c(t)+\hat\Gamma_c^{\dagger}(t)\hat{a}(t);\label{eq:Hcb}\\
\hat{H}_{mb}^{I}&=&[\hat{b}(t)+\hat{b}^{\dagger}(t)][\hat{\Gamma}_m(t)+\hat{\Gamma}_m^{\dagger}(t)].\label{eq:Hbm}
\end{eqnarray}
The system operator $\hat{a}(t)=e^{i\hat{H}_s t}\hat{a} e^{-i\hat{H}_s t}$ can be decomposed in terms of the eigenstates as 
\begin{equation}
\hat{a}(t)=\sum_{n, k, j}e^{-i\Delta_{k,j}^{(n)} t}A_{j, k}^{(n)}| n-1,j^{(n-1)}\rangle\, \langle n, k^{(n)}|,\label{eq:at}
\end{equation}
where the Franck-Condon factors $A_{j,k}^{(n)}=\sqrt{n}\langle j^{(n-1)}|k^{(n)}\rangle$ are finite for $j\ne k$, indicating that $\hat{a}(t)$ contains many phonon sidebands, and $\Delta_{k,j}^{(n)}=(\varepsilon_{n,k}-\varepsilon_{n-1,j})$. The operator $\hat{b}(t)=e^{i\hat{H}_s t}\hat{b} e^{-i\hat{H}_s t}$ can be simplified as 
\begin{equation}
\hat{b}(t)=e^{-i\omega_{m} t}(\hat{b}-\beta_{0}\hat{N}_c)+\beta_{0}\hat{N}_c\label{eq:btsimple}
\end{equation}
with $\hat{N}_c=\hat{a}^\dagger\hat{a}$ being the photon number operator. The operator $\hat{\Gamma}_{c}(t)$ ($\hat{\Gamma}_{m}(t)$) is the cavity (mechanical) bath operator with $\hat\Gamma_c(t)=\sum_{j}g_{cj}e^{-i\omega_{cj}t}\hat{c}_{cj}$ ($\hat\Gamma_m(t)=\sum_{j}g_{mj}e^{-i\omega_{mj}t}\hat{c}_{mj}$) in terms of the annihilation operator $\hat{c}_{cj}$ ($\hat{c}_{mj}$), frequency $\omega_{cj}$ ($\omega_{mj}$), and coupling constant $g_{cj}$ ($g_{mj}$) of the bath modes. With $\omega_{c}\gg\omega_{m}$, the cavity bath spectral density$J_c(\omega)=\sum_{j} |g_{cj}|^{2}\delta(\omega-\omega_{cj})$ can be assumed to be flat over the whole range of relevant phonon sidebands with $J_c(\omega_{c})=\kappa/2\pi$. We also assume that the mechanical bath spectral density $J_m(\omega)=\sum_{j} |g_{mj}|^{2}\delta(\omega-\omega_{mj})$ is of Ohmic form with $J_m(\omega)=(\gamma_{m}\omega/2\pi\omega_{m})$ and $\gamma_{m}$ being the mechanical damping rate. At high temperature, this spectral density corresponds to a white noise on the mechanical mode~\cite{Vitali2008}. 
\begin{figure}
\includegraphics[width=\columnwidth,clip]{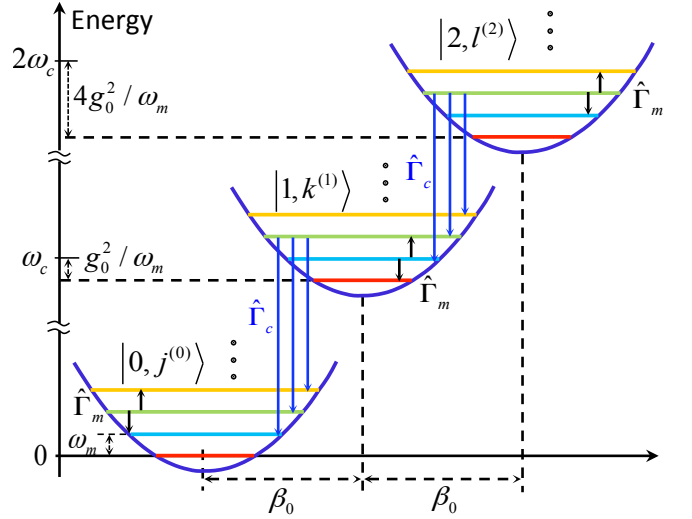}
\caption{(Color online) Schematic energy diagram and system-bath coupling of the optomechanical system (only the $0$-,  $1$- and $2$-photon subspaces shown). The arrows indicate transitions induced by cavity (blue) and mechanical (black) bath modes.}
\label{fig1}
\end{figure}
 
Under the Born-Markov and the rotating wave approximations (RWA), we then derive the full master equation of this system using the dressed-state operator decomposition given in Eqs.~(\ref{eq:at}) and (\ref{eq:btsimple}). The master equation in the Schr\"{o}dinger picture has the form
\begin{eqnarray} 
\frac{d\rho(t)}{dt}&=&-i[\hat{H}_s,\rho(t)]+\gamma_m(n_{th}+1)\mathcal{D}[\hat{b}-\beta_0\hat{N}_c]\rho(t)\nonumber\\
&+&\kappa\mathcal{D}[\hat{a}]\rho(t)+\gamma_m n_{th}\mathcal{D}[\hat{b}^\dagger-\beta_0\hat{N}_c]\rho(t)\nonumber \\
&+&4\gamma_{m}(k_{B}T/\omega_{m})\beta_{0}^{2}\mathcal{D}[\hat{N}_c]\rho(t),\label{eq:dsme} 
\end{eqnarray}
where $n_{th}$ is the thermal phonon occupation number at temperature $T$ and $\mathcal{D}[\hat{o}]\rho(t)$ is the Lindblad superoperator. Below we call this master equation the dressed-state master equation (DSME). The last term in this master equation is due to the low-frequency part of the mechanical noise~\cite{Vitali2008} and could induce dephasing between different photon number states. Detailed derivation of the DSME can be found in Appendix~\ref{smsec1}. In the limit of weak single-photon optomechanical coupling with $\beta_{0}\ll1$, the $\beta_{0}$-dependent terms in the DSME can be neglected. The DSME then becomes
\begin{eqnarray}
\frac{d\rho(t)}{dt}&=&-i[\hat{H}_s,\rho(t)]+\gamma_m(n_{th}+1)\mathcal{D}[\hat{b}]\rho(t)\nonumber \\
&+&\kappa \mathcal{D}[\hat{a}]\rho(t)+\gamma_m n_{th}\mathcal{D}[\hat{b}^\dagger]\rho(t),\label{eq:sme}
\end{eqnarray}
which has the familiar form of the SME often seen in the literature. 

Compared with SME, the extra terms in the DSME originate from the mechanical bath modes and the interaction between the cavity and the mechanical modes. This interaction results in the expression in Eq.~(\ref{eq:btsimple}). From Eq.~(\ref{eq:btsimple}) together with Eq.~(\ref{eq:Hbm}), we see that the mechanical resonator-bath coupling generates two physical processes: (i) the exchange of phonons between the system and bath modes in the shifted basis, which gives rise to the $\mathcal{D}[\hat{b}-\beta_0\hat{N}_c]$ and $\mathcal{D}[\hat{b}^{\dag}-\beta_0\hat{N}_c]$ terms in Eq.~(\ref{eq:dsme}); (ii) the shift of the mechanical displacement that depends on the photon number, which yields the last term in Eq.~(\ref{eq:dsme}). With $\beta_{0}\sim 1$ in the ultrastrong coupling regime, the extra terms can have a strong impact on the coherence and dynamics of the optomechanical system. 

\begin{figure}
\includegraphics[width=\columnwidth,clip]{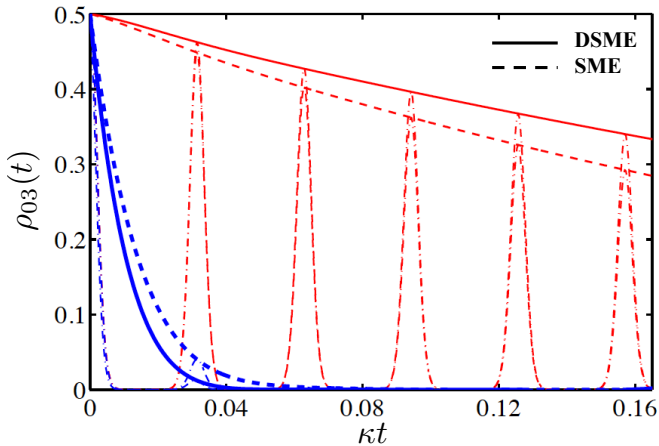}
\caption{(Color online) Time envelope of $|\rho_{03}(t)|$. Thin (red) envelopes are for $n_{th}=0$; thick (blue) envelopes are for $n_{th}=20$. The dotted curves are the actual time evolution of $|\rho_{03}(t)|$. Other parameters are $g_0=0.8\,\omega_{m}$, $\kappa=0.005\,\omega_{m}$ and $\gamma_{m}=0.00167\,\omega_{m}$.}
\label{fig2}
\end{figure}
\section{Cavity dephasing\label{sec3}} The dynamics of an optomechanical system governed by the DSME could be quite different from the dynamics governed by the SME. We first study the dephasing of cavity states. Consider the optomechanical system in an initial state $\vert \psi(0)\rangle=\frac{1}{\sqrt{2}}(\vert0\rangle_c+\vert3\rangle_c)\vert0\rangle_{m}$, with both the cavity and the mechanical modes in a pure state. We numerically simulate the time evolution of the density matrix of this system using the package in Ref.~\cite{toolbox}. We then calculate the off-diagonal matrix element $\rho_{03}(t)\equiv \vert _{c}\langle 0\vert \textrm{Tr}_{m}[\rho(t)]\vert 3 \rangle_{c}\vert$ of the density matrix $\rho(t)$, where $|0\rangle_{c}, \,|3\rangle_{c}$ are photon number states and $\textrm{Tr}_{m}$ is a trace operation over the mechanical mode. This matrix element directly reflects the coherence of the cavity mode. In Fig.~\ref{fig2}, $\rho_{03}(t)$ from the DSME as well as from the SME is plotted. At $n_{th}=0$ ($T=0$), the DSME result predicts stronger cavity coherence than that of the SME, with $\rho_{03}(t)$ decreasing at a slower rate with the DSME. However, at $n_{th}=20$, opposite behavior can be observed with $\rho_{03}(t)$ decreasing at a faster rate with the DSME than that with the SME. These results indicate that the dephasing of the cavity is strongly affected by the $\beta_{0}$-dependent terms in Eq.~(\ref{eq:dsme}) even at moderate thermal occupation number, and the SME is not sufficient to correctly describe the time evolution of this system. 

To explain the above result, we write the master equations in the interaction picture, which are given by Eq.~(\ref{eq:smdsme2Int}) and Eq.~(\ref{eq:smsme1int}) in Appendix~\ref{smsec1}. In the interaction picture, the bath-induced terms in the DSME are exactly the same as that of Eq.~(\ref{eq:dsme}), only with $\rho(t)$ replaced by the density matrix $\rho^{I}(t)$ in the interaction picture. Whereas in the SME, with all other terms staying the same as that in the DSME, the $\mathcal{D}[\hat{N}_c]$ term has a different coefficient: $\gamma_{m}(2n_{th}+1)\beta_0^2$. Hence at $n_{th}=0$ ($T=0$), the SME has one more term than the DSME: $\gamma_{m}\beta_{0}^{2}\mathcal{D}[\hat{N}_c]\rho^{I}(t)$, which explains the slower dephasing predicted by the DSME. At $n_{th}=20$ (finite $T$), the coefficient of the $\mathcal{D}[\hat{N}_c]$ term in the DSME becomes larger than that in the SME, which predicts faster dephasing for the DSME.

The time evolution of the photon number average, in contrast, is not affected by the $\beta_{0}$-dependent terms in the master equation. It can be shown that with DSME, $\langle\hat{N}_{c}(t)\rangle=\exp{(-\kappa t)}\langle\hat{N}_{c}(0)\rangle$, as given by Eq.~(\ref{eq:smNcave}) in Appendix~\ref{smsec1}, which is the usual photon exponential decay at a decay rate $\kappa$. 

\begin{figure}
\includegraphics[width=\columnwidth,clip]{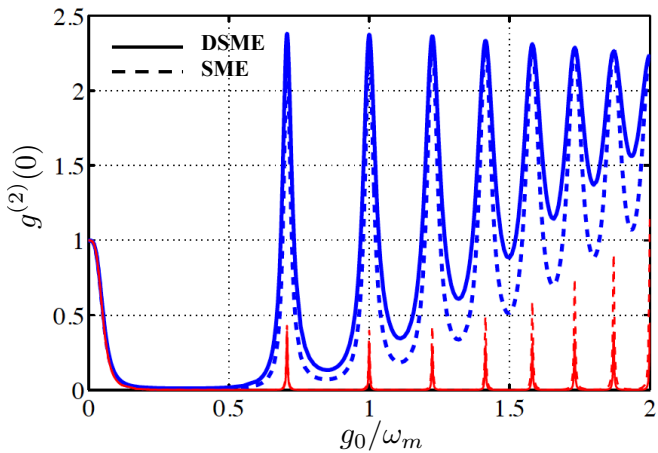}
\caption{(Color online) Photon correlation $g^{(2)}(0)$ versus $g_0/\omega_{m}$ at detuning $\Delta_0\equiv g_0^2/\omega_{m}$ and driving amplitude $E_{0}=0.1\kappa$. Thin (red) curves are for $n_{th}=0$; thick (blue) curves are for $n_{th}=10$. Other parameters are $\kappa=0.005\,\omega_{m}$ and $\gamma_{m}=0.0033\,\omega_{m}$.}
\label{fig3}
\end{figure}
\section{Second-order photon correlation\label{sec4}} Photon correlation can be strongly affected by the radiation-pressure interaction in an optomechanical system with ultrastrong coupling~\cite{Ultrastrong2, Kronwald2013, Liao2013}. The second-order photon correlation at equal times defined as $g^{(2)}(0) =\langle\hat{a}^\dagger\hat{a}^\dagger\hat{a}\hat{a}\rangle_{ss}/\langle\hat{a}^\dagger\hat{a}\rangle_{ss}^2$ is a widely used quantity to identify quantum features of a photon state such as antibunching. Here we study the behavior of $g^{(2)}(0)$ of an optomechanical system governed by the DSME and the SME. The system is under a weak driving on the cavity mode. With the driving, the Hamiltonian $\hat{H}_s$ in Eq.~(\ref{eq:dsme}) needs to be replaced by $\hat{H}_{s}^{\prime}=\hat{H}_s+E_{0}\left(\hat{a} e^{i\omega_{d}t}+\hat{a}^\dagger e^{-i\omega_d t}\right)$, where $E_{0}$ ($\omega_d$) is the amplitude (frequency) of the driving field. In our numerical calculation~\cite{toolbox}, we choose the detuning of the driving field $\Delta_0\equiv\omega_c-\omega_d$ to be at the single-photon resonance with $\Delta_0= g_0^2/\omega_{m}$, i.e., the driving field can resonantly excite the transition between the ground state and the state $|1,0^{(1)}\rangle$~\cite{Ultrastrong1, Ultrastrong2}. We derive the photon correlation by solving the steady state of the master equations. 

The photon correlation $g^{(2)}(0)$ is plotted in Fig.~\ref{fig3} as a function of the dimensionless constant $\beta_{0}=g_{0}/\omega_{m}$. Similar to that in previous works~\cite{Ultrastrong2, Liao2013}, $g^{(2)}(0)$ demonstrates oscillating behavior with peak positions at $\beta_{0}=\sqrt{k/2}$ for integer number $k$. These peaks correspond to two-photon resonances at given phonon sidebands. At $n_{th}=0$, the result with the DSME gives smaller $g^{(2)}(0)$ values and indicates more quantumness in the photon state than that with the SME. On the other hand, at $n_{th}=10$, $g^{(2)}(0)$ from the DSME is always larger than that from the SME, indicating less antibunching and weaker photon blockade. In particular, in the vicinity of $\beta_{0}=1.7$ and several other values, the SME gives $g^{(2)}(0)<1$; while the DSME gives an opposite result of $g^{(2)}(0)>1$, which shows that photon blockade does not occur. These numerical results can be explained by our previous analysis of the master equations in the interaction picture and also agree with the results for cavity dephasing. Our results imply that in the ultrastrong regime, the second-order photon correlation depends  sensitively on the coupling $\beta_{0}$ and could be strongly affected by the $\mathcal{D}[\hat{N}_c]$ term in the DSME. 

\section{Two-cavity entanglement\label{sec5}} Consider an optomechanical system made of two cavity modes coupling to a common mechanical resonator with the total radiation-pressure interaction $\hat{H}_{int}=-\sum_{i} g_{i} \hat{a}_{i}^{\dag}\hat{a}_{i}(\hat{b}+\hat{b}^{\dag})$, where $g_{i}$'s are the coupling constants and $\hat{a}_{i}$'s are the annihilation operators for the cavity modes with $i=1,2$. Here we study the entanglement between the two cavity modes. The DSME for this system can be derived as 
\begin{eqnarray} 
\frac{d\rho(t)}{dt}&=&-i[\hat{H}_s,\rho(t)]+\gamma_m(n_{th}+1)\mathcal{D}[\hat{b}-\hat{N}_t]\rho(t) \nonumber \\
&+&\sum_{i}\kappa_{i}\mathcal{D}[\hat{a}_{i}]\rho(t)+\gamma_m n_{th}\mathcal{D}[\hat{b}^\dagger-\hat{N}_t]\rho(t)\nonumber \\
&+&4\gamma_{m}(k_{B}T/\omega_{m})\mathcal{D}[\hat{N}_t]\rho(t),\label{eq:dsme2} 
\end{eqnarray}
where $\hat{H}_s$ is the total Hamiltonian with the interaction $\hat{H}_{int}$ given above, $\kappa_{i}$ is the damping rate of each cavity mode, and $\hat{N}_{t}=\beta_1\hat{N}_{c1}+\beta_2\hat{N}_{c2}$ with $\beta_i=g_i/\omega_m$ and $\hat{N}_{ci}=\hat{a}_{i}^{\dag}\hat{a}_{i}$. The difference between Eq.~(\ref{eq:dsme2}) and Eq.~(\ref{eq:dsme}) is that the $\hat{N}_t$ terms in the above master equation contain contributions from both cavities. Details of the derivation are presented in Appendix~\ref{smsec2}.
\begin{figure}
\includegraphics[width=\columnwidth,clip]{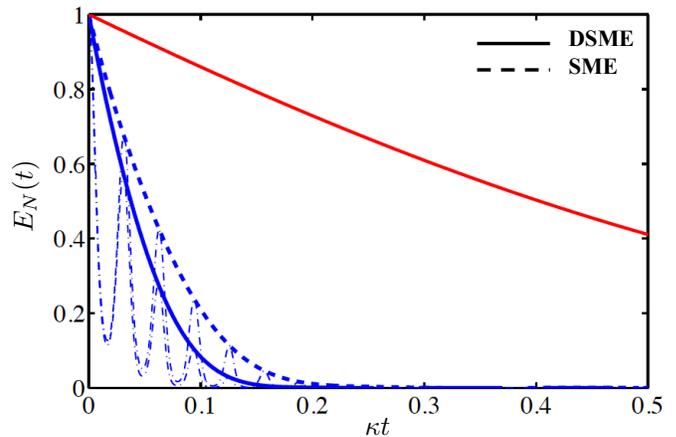}
\caption{(Color online) Time dependence of $E_{N}(t)$. Red curve is for $\beta_{1,2}=1.5$. Thick (blue) curves are envelopes of the time dependence (dotted curves) for $\beta_1=1.5$ and $\beta_2=0.5$. Other parameters are $\kappa_{1,2}=0.005\,\omega_{m}$, $\gamma=0.00167\,\omega_{m}$ and $n_{th}=20$.}
\label{fig4}
\end{figure}

We study the time dependence of the entanglement between the two cavity modes using the master equations. The system starts with an initial state $\vert \varphi(0)\rangle=\frac{1}{\sqrt{2}}[(\vert0\rangle_{c1}\vert1\rangle_{c2}+\vert1\rangle_{c1}\vert0\rangle_{c2})\vert0\rangle_{m}]$ with the cavities in a fully entangled state. We characterize the entanglement with the logarithmic negativity~\cite{entanglement1, entanglement2}: $E_{N}(t)=\log_{2}\| (\textrm{Tr}_{m}[\rho(t)])^{T_{A}} \|$, where the superscript ${T_{A}}$ denotes the partial transpose of the reduced density matrix $\textrm{Tr}_{m}[\rho(t)]$ and $\|\hat{o}\|$ denotes the trace norm of the matrix $\hat{o}$. The logarithmic negativity $E_{N}(t)$ is plotted in Fig.~\ref{fig4}. For equal coupling strength $\beta_{1,2}=1.5$, the results from the DSME and from the SME are exactly the same and without oscillations in the amplitudes. This is because $\hat{N}_t=\beta_1(\hat{N}_{c1}+\hat{N}_{c2})$ at equal coupling, proportional to the total photon number in the cavities; and our initial state is in a superposition of two states $\vert0\rangle_{c1}\vert1\rangle_{c2}$ and $\vert1\rangle_{c1}\vert0 \rangle_{c2}$, which have equal total photon number. Hence, the $\hat{N}_{t}$-dependent terms in the master equations generate equal phase fluctuations on these two states, and induce no extra dephasing in this special case. However, when the couplings are different, e.g., for $\beta_1=1.5$ and $\beta_2=0.5$, the DSME and the SME give different results. At $n_{th}=20$, $E_{N}(t)$ derived from the DSME decays faster than that from the SME, similar to the behavior of cavity dephasing shown in Fig.~\ref{fig2}, due to the larger $\mathcal{D}[\hat{N}_t]$ terms in the DSME. This indicates that the mechanical noise is transferred to the cavity modes via the optomechanical coupling and degrades the entanglement. Note that although the time envelopes in Fig.~\ref{fig2} and Fig.~\ref{fig4} all show exponential decay, their time scales and detailed behaviors are quite different. The similarity in the time envelops is due to the forms of the dissipative terms in the master equations, which induce this generic behavior in both cavity dephasing and entanglement. 

\section{Conclusions\label{sec6}} To summarize, we study the quantum coherence in an optomechanical system in the ultrastrong coupling regime with a dressed-state master equation approach. Compared with the standard approach,  our master equation takes into account the modification of the eigenstates due to the optomechanical coupling between the cavity and the mechanical modes, and predicts different behaviors in cavity dephasing, second-order photon correlation and two cavity entanglement. Our results show that ultrastrong light-matter interaction can play a significant role in the open system dynamics of an optomechanical system. This work could be useful for future studies of nonlinear optical effects in optomechanical systems. 

\section*{Acknowledgments} DH and LT are supported by the DARPA ORCHID program through AFOSR, the National Science Foundation under Award No. NSF-DMR-0956064, and the NSF-COINS program under Grants No. NSF EEC-0832819. JQL is supported by the JSPS Foreign Postdoctoral Fellowship under No.~P12503. HSG acknowledges support from the MOST in Taiwan under Grant No.~100-2112-M-002-003-MY3 and No.~103-2112-M-002-003-MY3,  from the National Taiwan University under Grants No.~NTU-ERP-103R891400 and No.~NTU-ERP-103R891402, and from the focus group program of the National Center for Theoretical Sciences, Taiwan.

\appendix
\section{DSME for single cavity system \label{smsec1}}
In this section, we present details of the derivation of the DSME given in Eq.~(\ref{eq:dsme}). The coupling between the system and the bath modes can be described by the Hamiltonian $\hat{H}_{b}^{I}=\hat{H}_{cb}^{I}+\hat{H}_{mb}^{I}$ in the interaction picture with the cavity-bath coupling $\hat{H}_{cb}^{I}$ given by Eq.~(\ref{eq:Hcb}) and the mechanical mode-bath coupling $\hat{H}_{mb}^{I}$ given by Eq.~(\ref{eq:Hbm}), respectively. Under the Born-Markov approximation, the master equation for the reduced density matrix $\rho^{I}(t)$ of the optomechanical system in the interaction picture can be derived as~\cite{qopticsbook, dePonte2004, BLHu2008, TianPRB2011}
\begin{equation}
\frac{d\rho^{I}(t)}{dt}=-\int_0^\infty ds \textrm{Tr}_{b}[\hat{H}_b^{I}(t),\,[\hat{H}_b^{I}(t-s),\,\rho^{I}(t)\otimes \rho_c\otimes\rho_{m}]], \label{eq:smrhoI}
\end{equation}
where $\textrm{Tr}_{b}$ denotes the trace operation over the bath modes and $\rho_c$ ($\rho_m$) is the density matrix of the cavity (mechanical) bath modes in their thermal state. As the cavity bath and the mechanical bath are independent from each other, the above master equation can be written as 
\begin{equation}
\frac{d\rho^{I}(t)}{dt}=\mathcal{L}_c^{I}\rho^{I}(t)+\mathcal{L}_m^{I}\rho^{I}(t),\label{eq:smcalL1}
\end{equation}
where $\mathcal{L}_{c}^{I}$ and $\mathcal{L}_{m}^{I}$ are superoperators acting on the density matrix of the system. By applying the rotating wave approximation (RWA) to remove fast oscillating terms such as the $e^{\pm 2i \omega_{c}t}$ terms, the cavity bath contribution becomes
\begin{eqnarray}
\mathcal{L}_c^{I}\rho^{I}(t) &=&\int_0^\infty ds \mathcal{R}_{-}(s) \hat {a}^\dagger(t-s)\rho^{I}(t)\hat{a}(t)  \nonumber \\
&-&\int_0^\infty ds \mathcal{R}_{-}(s) \hat{a}(t)\hat{a}^\dagger(t-s)\rho^{I}(t) \nonumber \\
&+& \int_0^\infty ds \mathcal{R}_{+}(s)\hat{a}(t-s)\rho^{I}(t)\hat{a}^\dagger(t) \nonumber \\
&-& \int_0^\infty ds \mathcal{R}_{+}(s)\hat{a}^\dagger(t)\hat{a}(t-s)\rho^{I}(t) \nonumber \\
&+& h.c. \label{eq:smLc}
\end{eqnarray}
with bath correlation functions defined as
\begin{eqnarray}
\mathcal{R}_{-}(s)&=&\textrm{Tr}_{b}\left[\hat{\Gamma}_c^\dagger(t)\hat{\Gamma}_c(t-s)\rho_c\right], \nonumber\\
\mathcal{R}_{+}(s)&=&\textrm{Tr}_{b}\left[\hat{\Gamma}_c(t)\hat{\Gamma}_c^\dagger(t-s)\rho_c\right]. \label{eq:smRpm}
\end{eqnarray}
For the mechanical bath, we have 
\begin{eqnarray}
\mathcal{L}_{m}^{I}\rho^{I}(t) &=& \int_0^\infty ds \mathcal{R}_m(s) \hat{X}(t-s)\rho^{I}(t)\hat{X}(t) \nonumber \\
&-& \int_0^\infty ds \mathcal{R}_m(s)\hat{X}(t)\hat{X}(t-s)\rho^{I}(t)\nonumber\\
&+& h.c.\label{eq:smLm}
\end{eqnarray}
with the time-dependent operators 
\begin{eqnarray}
\hat{X}(t)&=&\hat{b}(t)+\hat{b}^\dagger(t),\nonumber\\
\hat{X}_{\Gamma}(t)&=&\hat{\Gamma}_m(t)+\hat{\Gamma}_{m}^\dagger(t)\label{eq:smXm}
\end{eqnarray}
and the correlation function for the mechanical bath
\begin{equation}
\mathcal{R}_m(s)=\textrm{Tr}_{b}\left[\hat{X}_{\Gamma}(t)\hat{X}_{\Gamma}(t-s)\rho_m\right].\label{eq:smRms}
\end{equation}
Below we derive the contributions of the cavity and the mechanical bath modes respectively. 

\subsection{Cavity bath contribution\label{smsubsec1a}}
We first write down the time-dependent operator $\hat{a}(t)$. Define the operators 
\begin{equation}
\hat{A}_{j,k}^{(n)}=\sqrt{n}\langle j^{(n-1)}|k^{(n)}\rangle| n-1, j^{(n-1)}\rangle\, \langle n, k^{(n)}|\label{eq:smAjkn}
\end{equation}
and the energy separations $\Delta_{k,j}^{(n)}=(\varepsilon_{n,k}-\varepsilon_{n-1,j})$ in terms of the eigenenergies  $\varepsilon_{n,k}$. It can be shown that $\Delta_{k,j}^{(n)} =\omega_c+(k-j)\omega_m+(1-2n)g_0^2/\omega_m$, including phonon sidebands $(k-j)\omega_{m}$. We then have 
\begin{equation}
\hat{a}(t)=\sum_{n, k, j}e^{-i\Delta_{k,j}^{(n)} t}\hat{A}_{j, k}^{(n)}.\label{eq:smat}
\end{equation}
The cavity bath contribution to the DSME can be derived from Eq.~(\ref{eq:smLc}). With $\hat\Gamma_c(t)=\sum_{j}g_{cj}e^{-i\omega_{cj} t}\hat{c}_{cj}$, 
\begin{eqnarray}
\mathcal{R}_{-}(s)&=&\sum_{j}|g_{cj}|^{2} n(\omega_{cj},T)e^{i\omega_{cj} s},\nonumber\\
\mathcal{R}_{+}(s)&=&\sum_{j}|g_{cj}|^{2} [n(\omega_{cj},T)+1]e^{-i\omega_{cj} s},\label{eq:smRpmbath}
\end{eqnarray}
where $n(\omega_{cj},T)$ is the average occupation number of the corresponding bath mode. Because $\omega_{c}\gg\omega_{m}$, we assume that the cavity bath spectral density defined as $J_c(\omega)=\sum_j |g_{cj}|^{2}\delta(\omega-\omega_{cj})$ is slow-varying near $\omega=\omega_c$, and can thus be written as $J_c(\omega)\equiv \kappa/2\pi$ in the full range of the phonon sidebands. Hence,
\begin{eqnarray}
\int_0^\infty dse^{i\Delta_{k,j}^{(n)}\cdot s}\mathcal{R}_{+}(s) \approx \frac{\kappa}{2}[n(\omega_c,T)+1] \approx \frac{\kappa}{2},&&\nonumber\\
\int_0^\infty dse^{-i\Delta_{k,j}^{(n)} \cdot  s}\mathcal{R}_{-}(s)\approx \frac{\kappa}{2}n(\omega_c,T)\approx 0,&&\label{eq:smRpmintg}
\end{eqnarray}
where the thermal photon number at the cavity frequency $n(\omega_c,T)\approx 0$. The cavity bath contribution is hence
\begin{eqnarray}
\mathcal{L}_c^{I}\rho^{I}(t) &=& \frac{\kappa}{2} \sum_{k,j,n,l,i,r} \left\{ 2[e^{-i\Delta_{k,j}^{(n)} t}\hat{A}_{j, k}^{(n)}]\rho^{I}(t)[e^{i\Delta_{l,i}^{(r)} t}\hat{A}_{i, l}^{(r)\dag}] \right. \nonumber\\&-&[e^{i\Delta_{l,i}^{(r)} t}\hat{A}_{i, l}^{(r)\dag}][e^{-i\Delta_{k,j}^{(n)} t}\hat{A}_{j, k}^{(n)}]\rho^{I}(t)  \nonumber \\
&-& \left.\rho^{I}(t)[e^{i\Delta_{l,i}^{(r)}t}\hat{A}_{i, l}^{(r)\dag}][e^{-i\Delta_{k,j}^{(n)} t}\hat{A}_{j, k}^{(n)}] \right\},\label{eq:smLcfull}
\end{eqnarray}
which is simply $\mathcal{L}_c^{I}\rho^{I}(t) =\kappa \mathcal{D}[\hat{a}(t)]\rho^{I}(t)$. Here $\mathcal{D}[\hat{o}]\rho(t)=\frac{1}{2}[2\hat{o} \rho(t)\hat{o}^\dagger-\hat{o}^\dagger\hat{o}\rho(t)-\rho(t)\hat{o}^\dagger\hat{o}]$ is the Lindblad superoperator for operator $\hat{o}$. Under the RWA, the fast oscillating terms in this expression can be omitted from the above equation. 

By transforming Eq.~(\ref{eq:smLcfull}) to the Schr\"{o}dinger picture, the cavity bath contribution can be simplified as
\begin{equation}
\mathcal{L}_c \rho(t)=\kappa \mathcal{D}[\hat{a}]\rho(t),\label{eq:smLcSchroe}
\end{equation}
where $\mathcal{L}_c$ is a superoperator acting on the density matrix of the system modes in the Schr\"{o}dinger picture $\rho(t)=e^{-i\hat{H}_s t}\rho^{I}(t)e^{i\hat{H}_s t}$. The time-dependent factors in this superoperator are cancelled due to the transformation $e^{-i\hat{H}_s t}$. Eq.~(\ref{eq:smLcSchroe}) has exactly the same form as the cavity bath contribution in a standard master equation.

\subsection{Mechanical bath contribution\label{smsubsec1b}}
The time-dependent operator $\hat{b}(t)$ can be decomposed in the eigenbasis as
\begin{eqnarray}
\hat{b}(t)&=&\sum_{n,j}\left[ \sqrt{j} e\!^{-i\omega_m t} |n, (j-1)^{(n)}\rangle\langle n, j^{(n)}| \right.\nonumber\\
&+&\left. \beta_0 n |n, j^{(n)}\rangle\langle n,j^{(n)}|\right],\label{eq:smbt}
\end{eqnarray}
which can be simplified to give Eq.~(\ref{eq:btsimple}). Using the expression $\hat\Gamma_m(t)=\sum_{j}g_{mj}e^{-i\omega_{mj} t}\hat{c}_{mj}$, we derive the correlation function $\mathcal{R}_m(s)$ defined in Eq.~(\ref{eq:smRms}) as
\begin{eqnarray}
\mathcal{R}_{m}(s)&=&\sum_{j}|g_{mj}|^{2} n(\omega_{mj},T) e^{i\omega_{mj} s} \nonumber \\
&+& \sum_{j}|g_{mj}|^{2} [n(\omega_{mj},T)+1] e^{-i\omega_{mj} s},\label{eq:smRmbath}
\end{eqnarray}
where $n(\omega_{mj},T)$ is the thermal occupation number of bath mode $\hat{c}_{mj}$. We assume that the spectral density of the mechanical bath $J_m(\omega)=\sum_j |g_{mj}|^{2} \delta(\omega-\omega_{mj})$ is Ohmic and takes the form of $J_m(\omega)= \frac{\gamma_{m}\omega}{2\pi\omega_{m}}$ in the continuum limit of bath frequency. Here $\gamma_m=2\pi J_m(\omega_{m})$ is the mechanical damping rate. Note that for an Ohmic spectral density, the correlation function in Eq.~(\ref{eq:smRmbath}) can be converted to the familiar form in Ref.~\cite{Vitali2008} with
\begin{equation}
\mathcal{R}_{m}(s)=\frac{\gamma_{m}}{2\omega_{m}}\int_{-\infty}^{\infty}
\frac{d\omega}{2\pi} \omega e^{-i\omega\cdot
s}\left[\coth{(\omega/2k_{B}T)}+1\right], \label{eq:smRmbathcont}
\end{equation}
where we have applied the relation $n(-\omega, T)=-[n(\omega, T)+1]$. Similar to the calculation for the cavity bath in Sec.~\ref{smsubsec1a}, we find
\begin{eqnarray}
\int_0^\infty dse^{i\omega_{m}\cdot s}\mathcal{R}_{m}(s)&=& \frac{\gamma_{m}}{2}(n_{th}+1),\nonumber\\
\int_0^\infty dse^{-i\omega_{m}\cdot s}\mathcal{R}_{m}(s)&=& \frac{\gamma_{m}}{2}n_{th},\nonumber\\
\int_0^\infty ds\mathcal{R}_{m}(s)&=&\frac{\gamma_{m}}{2}\left(\frac{k_{B}T}{\omega_{m}}\right), \label{eq:smRmintg}
\end{eqnarray}
where $n_{th}\equiv n(\omega_m, T)$ is the thermal phonon number at the mechanical resonance. 

Using this result and applying the RWA to omit the fast oscillating terms, we derive the mechanical bath contribution to the DSME: 
\begin{eqnarray}
\mathcal{L}_{m}^{I}\rho^{I}(t)&=&\gamma_{m}(n_{th}+1)\mathcal{D}[\hat{b}-\beta_{0}\hat{N}_c]\rho^{I}(t) \nonumber\\
&+&\gamma_{m}n_{th}\mathcal{D}[\hat{b}^{\dag}-\beta_{0}\hat{N}_c]\rho^{I}(t)\nonumber \\
&+&4\gamma_{m}\left(\frac{k_{B}T}{\omega_{m}}\right)\beta_{0}^{2}\mathcal{D}[\hat{N}_c]\rho^{I}(t).\label{eq:smLmsimple}
\end{eqnarray}
With Eq.~(\ref{eq:btsimple}), $e^{-i\hat{H}_s t}(\hat{b}-\beta_{0}\hat{N}_c)e^{i\hat{H}_s t}=e^{i\omega_{m} t}(\hat{b}-\beta_{0}\hat{N}_c)$. The mechanical bath contribution in the Schr\"{o}dinger picture $\mathcal{L}_{m}\rho(t)$ has exactly the same form as that of Eq.~(\ref{eq:smLmsimple}) with the replacement $\rho^{I}(t)\rightarrow \rho(t)$.

\subsection{Master equations\label{smsubsec1c}}
Here we summarize the equations derived in the previous subsections. In the Schr\"odinger picture, the DSME has the form
\begin{equation}
\frac{d\rho(t)}{dt}=-i[\hat{H}_s,\rho(t)] +\mathcal{L}_{c}\rho(t)+ \mathcal{L}_{m}\rho(t)\label{eq:smdsme1}
\end{equation}
with $\mathcal{L}_{c}\rho(t)$ and $\mathcal{L}_{m}\rho(t)$ given by Eqs.~(\ref{eq:smLcSchroe}) and (\ref{eq:smLmsimple}), respectively. Written explicitly in terms of the system operators, we obtain the master equation given by Eq.~(\ref{eq:dsme}). In the interaction picture, the DSME becomes
\begin{eqnarray}
\frac{d\rho^{I}(t)}{dt}&=&\kappa \mathcal{D}[\hat{a}(t)]\rho^{I}(t)\nonumber \\
&+& \gamma_m(n_{th}+1)\mathcal{D}[\hat{b}-\beta_0\hat{N}_c]\rho^{I}(t)\nonumber\\
&+&\gamma_m n_{th}\mathcal{D}[\hat{b}^\dagger-\beta_0\hat{N}_c]\rho^{I}(t)\nonumber\\
&+&4\gamma_{m}\left(\frac{k_{B}T}{\omega_{m}}\right)\beta_{0}^{2}\mathcal{D}[\hat{N}_c]\rho(t),\label{eq:smdsme2Int}
\end{eqnarray}
which contains fast oscillating terms with frequency $O(\omega_{m})$ generated by the phonon sidebands. These terms can be omitted under the RWA. 

The SME in the Schr\"{o}dinger picture, often seen in the literature, is given by Eq.~(\ref{eq:sme}). Applying the transformation $\rho^I(t)=e^{i\hat{H}_s t}\rho(t)e^{-i\hat{H}_s t}$ and omitting the fast oscillating terms including $e^{\pm i\omega_{m}t}$, the SME in the interaction picture becomes
\begin{eqnarray}
\frac{d\rho^I(t)}{dt}&=&\kappa \mathcal{D}[\hat{a}(t)]\rho^I(t)\nonumber\\
&+& \gamma_m(n_{th}+1)\mathcal{D}[\hat{b}-\beta_0\hat{N}_c]\rho^I(t) \nonumber\\
&+&\gamma_m n_{th}\mathcal{D}[\hat{b}^\dagger-\beta_0\hat{N}_c]\rho^I(t) \nonumber\\
&+&\gamma_{m}(2n_{th}+1)\beta_0^2\mathcal{D}[\hat{N}_c]\rho^{I}(t).\label{eq:smsme1int}
\end{eqnarray}
Note that we have used Eq.~(\ref{eq:btsimple}) and the RWA in deriving this master equation. The difference between Eq.~(\ref{eq:smdsme2Int}) and Eq.~(\ref{eq:smsme1int}) is in the last term of the master equation, which corresponds to photon dephasing. The difference is proportional to $\gamma_m \left[4\left(\frac{k_{B}T}{\omega_{m}}\right)-2n_{th}- 1\right]\beta_0^2$, and is originated from the mechanical bath modes. Because of the strong coupling between the cavity and the mechanical modes, the mechanical noise is transferred to the cavity mode and induces photon dephasing. At high temperature with $k_{B}T\gg\omega_{m}$, the DSME predicts more serious dephasing than the SME. Whereas at low temperature, the DSME in Eq.~(\ref{eq:smdsme2Int}) predicts slower dephasing than the SME. We want to note that the master equations here are all based on the bath correlation function given by Eq.~(\ref{eq:smRmbathcont}), which corresponds  to a white noise spectrum on the mechanical mode at high temperature. 

\subsection{Analytical solutions of operator averages\label{smsubsec1d}}
With the DSME given above, the time evolution of some operators can be solved analytically. For the photon number operator $\hat{N}_c$, 
\begin{equation}
 \frac{d \langle\hat{N}_c\rangle}{dt}=-\kappa\langle\hat{N}_c\rangle,\label{eq:smNcave}
\end{equation}
which yields the solution $\langle\hat{N}_c(t)\rangle = e^{-\kappa t}\langle\hat{N}_c(0)\rangle$. This result is the exactly same as the time evolution given by the SME, i.e., the dynamics of the photon number operator is not affected by our approach. This is because $\hat{N}_c$ commutes with both $\hat{H}_s$ and the extra dephasing term (the last term) in the DSME.

Similarly, for the annihilation operator of the mechanical mode $\hat{b}$, 
\begin{equation}
\frac{d \langle\hat{b}\rangle}{dt}=-i\omega_m(\langle\hat{b}\rangle-\beta_{0}\langle\hat {N}_c\rangle)-\frac{\gamma_m}{2}(\langle\hat{b}\rangle-\beta_0 \langle\hat{N}_c\rangle),\label{eq:smbave}
\end{equation}
which depends on the photon number average $\langle\hat{N}_c\rangle$. Combining Eq.~(\ref{eq:smNcave}) and Eq.~(\ref{eq:smbave}), we derive 
\begin{eqnarray}
\langle\hat{b}(t)\rangle &=&  e^{-i\omega_m t-\frac{\gamma_m}{2}t}\langle\hat{b}(0)\rangle \label{eq:smbave}\\
&+& \frac{ig_0+\beta_0\gamma_m/2}{i\omega_m+\gamma_m/2 -\kappa}\left(e^{-\kappa t}-e^{-i\omega_m t-\frac{\gamma_m}{2} t}\right)\langle\hat{N}_c(0)\rangle, \nonumber 
\end{eqnarray}
which depends on the initial cavity photon number, but is independent of the thermal temperature of the mechanical bath. 

\section{DSME for two cavity system~\label{smsec2}}
In this section, we derive the DSME given by Eq.~(\ref{eq:dsme2}) for two cavity modes coupling to a common mechanical mode. The total Hamiltonian of this system can  be written as 
\begin{equation}
\hat{H}_{s}=\sum_{i=1,2} \omega_{ci}\hat{a}_i^{\dagger}\hat{a}_i+\omega_{m}\hat{b}^{\dagger}\hat{b}-\sum_{i=1,2} g_{i}\hat{a}_i^{\dagger}\hat{a}_i(\hat{b}^{\dagger}+\hat{b}),\label{eq:smHs}
\end{equation}
where $\hat{a}_{i}$ is the annihilation operator for the $i$th cavity mode, $\omega_{ci}$ is its frequency, and $g_{i}$ is the coupling constant between cavity $\hat{a}_{i}$ and the mechanical mode. The eigenstates of this Hamiltonian are
\begin{equation}
|n_{1}, n_{2}, k^{(n_1,n_2)}\rangle=|n_{1}\rangle_{c1} |n_{2}\rangle_{c2} e^{(\sum_i n_i\beta_i )(\hat{b}^\dagger-\hat{b})}|k\rangle_{m}\label{eq:smeigstate}
\end{equation}
with $\beta_i=g_{i}/\omega_{m}$. The corresponding eigenenergies are
\begin{equation}
\varepsilon_{n_1,n_2,k}=n_1 \omega_{c1}+n_2\omega_{c2}+k\omega_{m}-(n_1\beta_1+n_2\beta_2)^2\omega_{m}.\label{eq:smen2cav}
\end{equation}
To derive the DSME, we consider the time-dependent operators $\hat{b}(t)$ and $\hat{a}_{i}(t)$. For the mechanical mode, 
\begin{equation}
\hat{b}(t)=e^{i\hat{H}_s t}\hat{b} e^{-i\hat{H}_s t}=e^{-i\omega_{m} t}(\hat{b}-\hat{N}_t)+\hat{N}_t\label{eq:smbt2cav}
\end{equation}
with the effective number operator defined as 
\begin{equation}
\hat{N}_t=\beta_1\hat{a}_{1}^{\dag}\hat{a}_{1}+\beta_2\hat{a}_{2}^{\dag}\hat{a}_{2}.\label{eq:smnc}
\end{equation}
For the cavity mode, $\hat{a}_{i}(t)=e^{i\hat{H}_s t}\hat{a}_{i} e^{-i\hat{H}_s t}$, including many phonon sidebands. We use the same assumptions as that in Appendix~\ref{smsec1}, i.e., the cavity spectral density is smooth in the entire range of the phonon sidebands and the mechanical bath is Ohmic. By applying the same procedure as that in Appendix~\ref{smsec1}, the DSME in Eq.~(\ref{eq:dsme2}) can be derived.

\end{document}